\documentstyle[psfig]{mn}

\newif\ifAMStwofonts
\def\lapp{\ifmmode\stackrel{<}{_{\sim}}\else$\stackrel{<}{_{\sim}}$\fi}
\def\gapp{\ifmmode\stackrel{>}{_{\sim}}\else$\stackrel{>}{_{\sim}}$\fi}

\title[Vector alignment and the emission heights in pulsars]
{Evidence for alignment of the rotation and velocity vectors
in pulsars. II. Further data and emission heights}
\author[Johnston et al.]
{Simon Johnston$^1$, M. Kramer$^2$, A. Karastergiou$^3$, G. Hobbs$^1$,
\newauthor S. Ord$^4$ \& J. Wallman$^4$\\
$^1$Australia Telescope National Facility, CSIRO, P.O. Box 76, 
Epping, NSW 1710, Australia. \\
$^2$ University of Manchester, Jodrell Bank Observatory, Macclesfield, Cheshire, SK11 9DL.\\
$^3$ IRAM, 300 rue de la Piscine, Domaine Universitaire, Saint Martin d'Heres, France.\\
$^4$ School of Physics, University of Sydney, NSW 2006, Australia.\\
}
\date{\today}
\pagerange{\pageref{firstpage}--\pageref{lastpage}}
\pubyear{2007}
\begin{document}
\maketitle
\label{firstpage}

\begin{abstract}
We have conducted observations of 22 pulsars at frequencies of
0.7, 1.4 and 3.1~GHz and present their polarization profiles.
The observations were carried out for two main purposes.
First we compare the orientation
of the spin and velocity vectors to verify the proposed alignment of these
vectors by Johnston et al. (2005). We find, for the 14 pulsars for which we
were able to determine both vectors, that 7 are plausibly aligned,
a fraction which is lower than, but consistent with, earlier measurements.
Secondly, we use profiles obtained simultaneously
at widely spaced frequencies to compute the radio emission heights.
We find, similar to other workers in the field,
that radiation from the centre of the profile originates from lower
in the magnetosphere than the radiation from the outer parts of the profile.
\end{abstract}


\section{Introduction}
During the first forty years of observations of radio pulsars,
various models have been developed to describe the basic
characteristics of pulsar emission
(e.g. Rankin 1983; Lyne \& Manchester 1988\nocite{ran83,lm88}).
The application of these models to
large amounts of observational data has provided fundamental knowledge of the
emission mechanism and the geometry of the emitting regions of pulsars
(e.g. Ruderman \& Sutherland 1975; Dyks et al. 2004\nocite{rs75,drh04}).

An observational paradigm which has come under scrutiny recently is
the radius to frequency mapping idea. Here, the emission height
depends on the observing frequency with lower frequencies arriving
from higher altitudes in the stellar magnetosphere. Several methods
are available for computing emission heights; these are described in
detail most recently by Mitra \& Li (2004)\nocite{ml04}.
In this paper, armed with full polarization information, we will
examine heights derived through both the aberration-retardation (A/R) 
method and the geometrical method.
In the former, pioneered by Blaskiewicz, Cordes \& Wasserman
(1991)\nocite{bcw91} and later refined by Hibschman \& Arons
(2001)\nocite{ha01}, an offset between the inflexion point of the
position angle swing and the midpoint of (symmetrical) profiles is
used to compute the emission heights.  Gangadhara \& Gupta
(2001)\nocite{gg01} take this further by computing emission heights
from the offset of different components from the profile midpoint. In
the geometrical method the observed pulse width, coupled with
knowledge of the angle between the rotation and magnetic axes
($\alpha$) and the sightline impact parameter ($\beta$), allows a determination
of the cone opening angle and hence the emission height.

The present understanding from these recent studies is that the
central emission occurs from low down in the magnetosphere and that
the altitude of emission increases towards the outer edges of the beam.
If this is correct, then emission heights computed via the A/R method
(which measures the height of the central component)
should be systematically {\it lower} than those computed through
the geometric method (which measures the heights of the outer components).
A further crucial point is that the absolute emission heights
seem largely independent of pulsar period, at least
at frequencies near 1~GHz \cite{ran93}, and are 
generally no more than 1000~km (Blaskiewicz et al. 1991\nocite{bcw91}).
Finally, radius to frequency mapping still occurs in the sense that
the entire U-shaped emission region moves upwards at lower frequencies.
The functional dependence follows a law described by Thorsett (1991)
\nocite{tho91a} and later extended by Xilouris et al. (1996)\nocite{xkj+96}
and Mitra \& Rankin (2002)\nocite{mr02a}.

One way to determine emission heights is to obtain highly accurate
time alignment of profiles at multiple frequencies coupled with
knowledge of the dispersion measure (DM). Furthermore, absolute
determination of the position angles at different frequencies
is also crucial. In this paper we use the simultaneous nature
of the dual 0.7/3.1~GHz receiver at the Parkes radio telescope
coupled with our
previous calibration techniques to obtain both time domain
and position angle alignment.

We can also use these profiles to further our study of
the correlation between the velocity vectors and the spin
axes of pulsars. In Johnston et al. (2005)\nocite{jhv+05}
we showed conclusively that, in the majority of cases, the 
rotational axis and the velocity vector of pulsars are indeed aligned
thus settling an outstanding problem in pulsar physics.
Rankin (2007)\nocite{ran07} re-examined the Johnston et al. (2005)
sample and added a further 24 sources for which good proper motions
and polarization profiles were available. She concluded that the case for
alignment was indeed very strong.
These studies have instigated new examination of the physical causes
of such an alignment. The work of Wang, Lai \& Han (2006)\nocite{wlh06}
considered hydrodynamic and magneto-neutrino kicks. The kick mechanism itself 
was considered in more detail by Ng \& Romani (2007)\nocite{nr07}. Their model
predicts that slow moving pulsars may not be aligned, whereas fast
moving pulsars should have their spin and velocity vectors aligned.

The observational success of this part of the project depends on an 
accurate determination
of the position angle of the rotation axis on the sky, which we
denote as PA$_r$. The value of PA$_r$ can be obtained from PA$_0$,
the position angle of the polarization at the point of approach
of the observer's line of sight to the magnetic pole.
This can be determined in one of two ways.
First, in the rotating vector model (RVM; Radhakrishnan \& Cooke
1969, Komesaroff 1970\nocite{rc69a,kom70}),
the radiation is beamed along the field lines
and the plane of polarization is determined by the projected angle of the
magnetic field line as it sweeps past the line of sight. The PA as a
function of pulse longitude, $\phi$, can be expressed as
\begin{equation}
{\rm PA} = {\rm PA}_{0} + 
{\rm arctan} \left( \frac{{\rm sin}\alpha 
 \, {\rm sin}(\phi - \phi_0)}{{\rm sin}\zeta
 \, {\rm cos}\alpha - {\rm cos}\zeta
 \, {\rm sin}\alpha \, {\rm cos}(\phi - \phi_0)} \right)
\end{equation}
Here, $\alpha$ is the angle between the rotation axis and the magnetic
axis. Defining $\beta$ as the impact parameter
(the angle between the line of sight to the magnetic axis at closest
approach), then $\zeta=\alpha+\beta$.
$\phi_0$ is the corresponding pulse longitude at which the PA is
then PA$_{0}$. 

As discussed in Johnston et al. (2005), the added complication of
orthogonal mode emission must be taken into account when deriving
PA$_r$ from PA$_0$. In the case of some pulsars, such as the Vela pulsar,
PA$_r$ = PA$_0$ + 90\degr\ because the pulsar predominantly emits
radiation which is polarized perpendicular (rather than parellel) to
the magnetic field lines \cite{lm95}.

Even without the complication of orthogonal mode emission,
fits to the RVM are not always possible for a variety of
reasons. Determination of PA$_0$ then relies on clues obtained directly
from the profile information, such as symmetry, changing sign of
circular polarization and spectral index effects. Multi-frequency 
observations are extremely useful in this regard and we use the current
data and previously published information to help arrive at our conclusions.

Table~\ref{sources} lists the pulsars observed, none of which are binary or
millisecond pulsars. The first two columns list the
J and B names of the pulsars. Column 3 gives the logarithm of the
pulsar's characteristic age, $\tau_c = P/2\dot{P}$, with $P$ the
pulsar period and $\dot{P}$ the period derivative. Column 4 gives
the distance to the pulsar in kpc.  Columns 5 to 8 give
the proper motions in Right Ascension and Declination,
the reference for these values, and PA$_v$, the
position angle of the velocity vector on the sky measured
counter-clockwise from north.  In all cases the number in brackets indicates
the error on the last digit(s).
Proper motions for these pulsars were derived either using the timing
technique described in Hobbs et al. (2004)\nocite{hlk+04}
or through conventional VLBI techniques.

\section{Observations and data reduction}
The observations were carried out using the Parkes radio telescope
on two separate occasions.
In the initial session, on 2005 Oct 14 and 15
we used the H-OH receiver at a central frequency of 1.369~GHz with 
a bandwidth of 256~MHz.

In the second session which took place from 2005 Oct 24 to 26 we used a 
dual frequency receiver system capable of observing simultaneously in both 
the 50 and 10~cm bands.  We used central frequencies of 3.1~GHz with
a bandwidth of 512~MHz and 0.69~GHz with an effective 
bandwidth (after interference rejection) of 35~MHz.

All receivers have orthogonal linear feeds and also have a pulsed
calibration signal which can be injected at a position angle of
45\degr\ to the two feed probes.  A digital correlator
was used which subdivided the bandwidth into 1024 frequency channels
and provided all four Stokes' parameters. We also recorded 1024
phase bins per pulse period for each Stokes' parameter.

The pulsars were observed for 15-30 minutes on each occasion.
Prior to the observation of
the pulsar a 3-min observation of the pulsed calibration signal was
made.  The data were written to disk in FITS format for subsequent
off-line analysis.

Data analysis was carried out using the PSRCHIVE
software package \cite{hvm04} and the 
analyis and calibration were carried out in an identical fashion
to that described in detail in Johnston et al. (2005). Most importantly,
we are able to determine absolute position angles for the linearly
polarized radiation at all three of our observing frequencies.
In order to determine the absolute position angles at the pulsar, we
must make a frequency dependent correction for the rotation of the 
plane of polarization caused by the interstellar medium.

We determined this rotation measure (RM) in a bootstrap fashion. First,
we measure the PA difference across the 256~MHz of bandwidth at 1.4~GHz
and determine the RM to an accuracy of a few units. Because we have
absolute PAs at all three observing frequencies we are then able to
determine a highly accurate RM by comparing the PAs at 0.69 and 3.1~GHz.
The major uncertainity in the final value of RM comes from the fact
that the alignment between the 0.69 and 3.1~GHz profiles is not always clear
(see Karastergiou \& Johnston 2006\nocite{kj06}).
\begin{table*}
\caption{Proper motions, rotation measures and polarisation position
angles for a sample of 22 pulsars.}
\setlength{\tabcolsep}{3pt}
\begin{tabular}{llccrrrrrrr}
\hline & \vspace{-3mm} \\
& & \multicolumn{1}{c}{Age} & \multicolumn{1}{c}{Dist}
& \multicolumn{4}{c}{Proper Motion}
& \multicolumn{1}{c}{Rotation} & \multicolumn{1}{c}{Dispersion} &
\multicolumn{1}{c}{Poln} \\
\multicolumn{1}{c}{Jname} & \multicolumn{1}{c}{Bname} 
& \multicolumn{1}{c}{log($\tau_c$)} &
& \multicolumn{1}{c}{$\mu_\alpha$} & \multicolumn{1}{c}{$\mu_\delta$}
& \multicolumn{1}{c}{Ref} & \multicolumn{1}{c}{PA$_v$} 
& \multicolumn{1}{c}{Measure} & \multicolumn{1}{c}{Measure}
& \multicolumn{1}{c}{PA$_0$} \\
& & \multicolumn{1}{c}{(yr)} & \multicolumn{1}{c}{(kpc)} 
&\multicolumn{1}{c}{(mas~yr$^{-1}$)}
& \multicolumn{1}{c}{(mas~yr$^{-1}$)} & & \multicolumn{1}{c}{(deg)}
& \multicolumn{1}{c}{(rad~m$^{-2}$)} & \multicolumn{1}{c}{(cm$^{-3}$pc)}
& \multicolumn{1}{c}{(deg)} \\
\hline & \vspace{-3mm} \\
J0452$-$1759 & B0450$-$18 & 6.2 & 3.1 & 12(3) & 4(5) & 1 & 72(23) & 11.1(3) & 39.72(4) & 47(3)\\
J0528+2200 & B0525+21 & 6.2 & 2.3 & $-$20(19) & 7(9) & 4 & & $-$40.2(3) & 51.85(11) & 34(3)\\
J0543+2329 & B0540+23 & 5.4 & 3.5 & 19(7) & 12(8) & 4 & 58(20) & 2.7(6) & 77.55(2)\\
J0614+2229 & B0611+22 & 5.0 & 4.7 & $-$4(5) & $-$3(7) & 4 & & 66.0(3) & 96.95(10) & \\
J0659+1414 & B0656+14 & 5.0 & 0.3 & 44.1(6) & $-$2.4(3) & 3 & 93.1(4) & 23.0(3) & 13.66(22) & $-$86(2)\\
\vspace{-2mm} \\
J0738$-$4042 &  B0736$-$40  & 6.6 & 2.1$^{a}$& $-$14.0(12)& 13(2) & 2 & 313(5) & 12.5(6) & 160.97(1) & $-$21(2)\\
J0837+0610 &  B0834+06  & 6.5 & 0.7 & 2(5) & 51(3) & 1 & 2(5) & 26.5(6) & 13.04(10) & 18(5) \\
J0837$-$4135 & B0835$-$41 & 6.5 & 4.2 & $-$2.3(18)& $-$18(3)  & 2 & 187(6) & 145(1) & 147.21(1) & $-$84(5)\\
J1559$-$4438 & B1556$-$44 & 6.6 & 1.6 & 1(6) & 14(11) & 5 & & $-$5.0(6) & 55.92(1) & 71(3)\\
J1604$-$4909 & B1600$-$49 & 6.7 & 3.6 & $-$30(7) & $-$1(3) & 6 & 268(6) & 34(1) & 140.71(2) & $-$17(3)\\
\vspace{-2mm} \\
J1735$-$0724 & B1732$-$07 & 6.7 & 4.3 & $-$2.4(1.7) & 28(3) & 2 & 355(3) & 34.5(3) & 73.46(7) & 55(5)\\
J1752$-$2806 & B1749$-$28 & 6.0 & 1.5 & $-$4(6) & $-$5(5) & 5 & & 95.1(3) & 50.33(2) \\
J1801$-$2451 & B1757$-$24 & 4.2 & 4.6 & $<14$ & $-$$-$ & 7 & 270 & 603.0(5) & 291.08(16) & $-$55(5)\\
J1820$-$0427 & B1818$-$04 & 6.2 & 2.4 & $-$6.2(3) & 15.6(8) & 1 & 338(17) & 67.5(6) & 84.50(3) & 42(3)\\
J1825$-$0935 & B1822$-$09 & 5.4 & 1.0 & $-$13(11) & $-$9(5) & 5 & & 66.2(3) & 19.43(5) & 10(10) \\
\vspace{-2mm} \\
J1841$-$0425 &  B1838$-$04  & 5.7 & 5.2 & 1(3) & 5(10) & 1 & & 416.0(8) & 325.21(26) & 3(3)\\
J1850+1335 &  B1848+13  & 6.6 & 3.1 & $-$17(4) & $-$11(6) & 1 & 237(16) & 152.7(6) & 59.99(14) & $-$45(3)\\
J1915+1009 &  B1913+10  & 5.6 & 5.3 & 1.1(3.0) & $-$11(5) & 1 & 174(15) & 430.0(6) & 241.50(8) & 85(3)\\
J1916+0951 &  B1914+09  & 6.2 & 2.9 & $-$6.8(4.0) & $-$3.1(8.0) & 1 & & 97(1) & 60.95(11) \\
J1937+2544 &  B1935+25  & 6.7 & 2.8 & $-$10(3) & $-$12(3) & 1 & 220(9) & 26.0(3) & 53.27(9) & $-$9(5)\\
\vspace{-2mm} \\
J2048$-$1616 & B2045$-$16 & 6.5 & 0.6 & 117(5) & $-$5(5) & 5 & 92(2) & $-$10.0(3) & 11.84(4) & $-$13(5)\\
J2330$-$2005 & B2327$-$20 & 6.7 & 0.5 & 74.7(19) & 5(3) & 2 & 86(2) & 9.5(6) & 7.77(25) & 21(10)\\
\hline & \vspace{-3mm} \\
\end{tabular}

Notes:
a. distance from H{\sc I} measurements \cite{jkww96}.
Proper motion references:
1. updated parameters since Hobbs et al. (2005)\nocite{hllk05},
2. Brisken et al. (2003a)\nocite{bfg+03},
3. Brisken et al. (2003b)\nocite{btgg03},
4. Harrison et al. (1993)\nocite{hla93},
5. Fomalont et al. (1997)\nocite{fgm+97},
6. Bailes et al. (1990)\nocite{bmk+90b},
7. Blazek et al. (2006)\nocite{bgc+06}.
\label{sources}
\end{table*}

\subsection{DM determination and time alignment}
A new feature of this work is the simultaneous acquisition of
the 0.69 and 3.1~GHz profiles. Because the data are simultaneous the
majority of the offset between the arrival of the pulse profiles is due
to the different dispersion delays at the two frequencies.
Knowledge of a fiducial point on the pulse profile could then
be used to compute the DM.  Often this is achieved through a cross
correlation of the pulse profiles (e.g. Ahuja et al. 2005\nocite{agmk05}).
However, this would negate the effect
we are looking for viz the possible shift of the profiles at
different frequencies due to magnetospheric effects.
We therefore need an accurate DM in order to align the profiles.

This is achieved by using the 1.4~GHz data,
measuring the DM across the band under the assumption that 
the pulse profile does not change over that frequency interval.
We ignore the instrumental effects such as path length differences
between the two receiver systems; these effects should be no more than a few
$\mu$s.  Our technique yields a typical accuracy in the arrival
times significantly better than 1\degr\ of longitude, more than good
enough for our purposes.

\section{Results}
The measured values of interstellar RM are listed in column 9
of table~\ref{sources} with the errors indicated in brackets.
The error in RM is typically less than
0.5~rad~m$^{-2}$ for pulsars with dual frequency measurements.
This small error in RM
is crucial for determining absolute position angles.
The DM and its error, as determined from the 1.4~GHz profiles, are
listed in column 10.

In this section we first describe the polarization profiles in detail and
the determination of PA$_0$. Following this we then look at the
emission heights of various components in the pulse profile where
possible. Four pulsars are in common with the Rankin (2007) sample.
We discuss similarities and differences in our findings on these objects.

\subsection{Pulsar Profiles and Determination of PA$_0$}
The following paragraphs refer to Figures 1 through 6, where the polarization
profiles are shown, generally at 0.69 and 3.1~GHz but also at
1.4~GHz where appropriate.  The amplitude scale is normalized to 
the brightest component.  All comparisons of amplitude
are relative to the profile components at that specific frequency.

\noindent{\bf PSR~J0452$-$1759}: The profile of this pulsar has
complex frequency evolution. At the lowest frequency the profile is
triple peaked with components of roughly equal width. At 3.1~GHz, the
central component is significantly weaker and an extra component has
appeared on the trailing edge.  A fit to the RVM (after taking into account
the orthogonal mode jumps) allows the inflexion
point of the PA to be well determined, especially at the lowest
frequency. This inflexion point occurs later than the profile midpoint.
We choose the inflexion to be $\phi_0$ and then PA$_0$ is 47\degr.

\noindent{\bf PSR~J0528+2200}: This pulsar has a classic double profile
at all frequencies, with the profile narrowing at high frequencies.
The PA swing shows the perfect behaviour expected from the RVM (though
rarely seen!). The inflexion point of the RVM fit occurs exactly at
the symmetry point of the profile at both 0.7 and 3.1~GHz and hence
PA$_0$ is 34\degr.

\noindent{\bf PSR~J0543+2329}: The profile of this pulsar likely
reveals that we are only detecting emission from the leading edge of
the cone. There is little evolution with frequency except that the
linear polarization declines and the circular polarization increases.
Despite the high signal to noise of the PA measurements at both
frequencies, differences in the PA profile are evident.
An RVM fit to our profile at 0.69~GHz shows the inflexion point occurs
close to where the polarization data ends (some 24\degr\ after the midpoint).
This is a similar result to that in Blaskiewicz et al. (1991), whilst
Weisberg et al. (2004)\nocite{wck+04} show polarized emission extends out 
even further at 0.43~GHz.
Formally, however, these fits are very poorly constrained. Although we
believe there is good evidence that $\phi_0$ is much later than the
pulse peak, and, tantalisingly, that PA$_0$ is close to PA$_v$ at
late longitudes, we cannot in good faith assign PA$_0$ to this pulsar.

\noindent{\bf PSR~J0614+2229}: This pulsar has an age of less than
10$^5$~yr, and shows all the characteristics of a young pulsar
profile \cite{jw06}. There is little profile evolution
with frequency and the linearly polarized fraction remains high at
all frequencies. This is likely to be a component far from the
magnetic axis in both the longitudonal and radial directions. Like
the previous pulsar, $\phi_0$ will then be located significantly later than
the total intensity profile.
We therefore cannot assign PA$_0$ for this object.

\noindent{\bf PSR~J0659+1414}:  Although this pulsar has an age slightly
in excess of 10$^5$~yr, its profile is very similar to that of
PSR~J0614+2229 and other young pulsars. However, the polarization profile
is asymmetric at 3.1~GHz and the pulsar has become completely depolarized
at 8.4~GHz \cite{jkw06} unlike other young pulsars.
Rankin (2007) provides convincing evidence that the profile is leading
edge with the magnetic pole crossing at later longitudes.
This is backed up by the RVM fits in Everett \& Weisberg (2001)\nocite{ew01}
who show that $\phi_0$ occurs some 15\degr\ after the profile peak.
Our RVM fits give similar results for $\phi_0$ and at this longitude
PA$_0$ is then $-$86\degr.

\noindent{\bf PSR~J0738$-$4042}: Scattering measurements for this
pulsar at 327~MHz \cite{rmdm+97} indicate that the scattering time at
0.69~GHz is $\sim$4~ms (or 4\degr\ of longitude) and indeed the low
frequency profile shows indication of scatter broadening. In particular,
we attribute the difference in the PA profiles between the low and
high frequencies, especially at the trailing edge of the profile,
to scattering.
The frequency evolution of the profile indicates that the main peak occurs
near the magnetic pole crossing. The RVM fit to the profile at 1.4~GHz
(after accounting for two orthogonal mode jumps) is excellent and
$\phi_0$ corresponds to the profile peak. At this location PA$_0$ is
$-$21\degr. 

\noindent{\bf PSR~J0837+0610}: The profile of this pulsar would appear
to be a classic double but the frequency evolution is not as expected.
At frequencies below about 2~GHz the leading component dominates but
there is little difference between the 0.1 \cite{kl99b} and 1.4~GHz profiles.
However, at 3.1~GHz and 4.5~GHz \cite{hkk98}
the two components are equal in strength.
The profile also appears to widen slightly
as the frequency increases. The linear polarization is low and the
circular polarization is negative throughout.
Mitra \& Rankin (2002)\nocite{mr02a} classify the double as belonging
to an inner cone.
Although the PA swing is somewhat odd looking, it is very similar at
both 0.69 and 1.4~GHz in our data. Taking the symmetrical
midpoint of the profile to be $\phi_0$, then PA$_0$ is 18\degr.

\noindent{\bf PSR~J0837$-$4135}: The pulse profile shows a bright
central component, with outrider components beginning to become more
prominent at higher frequencies. The polarization behaviour is rather
complex with at least two PA jumps at 0.69~GHz, with a segment between
them that has a clear non-orthogonal offset. The steepest swing of
polarization appears to occur later than the profile midpoint by about
4\degr. After accounting for the mode jumps, a reasonably good RVM fit
can be obtained. We place $\phi_0$ at the inflexion point, which
occurs later than the profile midpoint and PA$_0$ is hence $-$84\degr.
This is in good agreement with the value obtained by Rankin (2007).

\noindent{\bf PSR~J1559$-$4438}: The pulse profile shows strong frequency
evolution. At low frequencies it consists of a single main profile
with a very weak leading component. At 3.1~GHz in contrast, the 
leading component has become stronger relative to the central component
and at least three new components can be seen on the trailing part of
the profile. The swing of PA at 0.69 and 1.4~GHz is well behaved apart
from two orthogonal mode jumps whereas at 3.1~GHz a strange U-shaped featured
is seen near the pulse centre.
An RVM fit to the 0.69 and 1.4~GHz profiles give similar results
with a small error bar on the location of the inflexion point which
occurs in advance of the low frequency component and the midpoint of
the 3.1~GHz profile.  Associating $\phi_0$ with the inflexion point
gives PA$_0$ of 71\degr.

\noindent{\bf PSR~J1604$-$4909}: At low frequencies a small component
is seen leading a large, possibly blended component. The circular
polarization swings from positive to negative near the centre of
the dominant component. At high frequencies, the leading component is
relatively larger and there is a hint of a trailing component.
The dominant component is narrower and the circular polarization no
longer changes sign. The PA swing is steep though confused in the
central component. We note that our derived value of RM, 34~radm$^{-2}$,
is significantly different from the previous value in the literature.
If $\phi_0$ is at the pulse peak then PA$_0$ is $-$17\degr.

\noindent{\bf PSR~J1735$-$0724}: The profile shows a strong central
component flanked by two outriders which are more prominent at high
frequencies. The linear polarization is rather weak and the PA swing
shows a hump-like feature over the central component likely indicating
an orthogonal mode jump, which is confusing the true nature of the RVM
swing. However, the peak of the main component clearly indicates
$\phi_0$ and we estimate that PA$_0$ is 55\degr\ there.

\noindent{\bf PSR~J1752$-$2806}: The profile of this pulsar is
characterised by a single peak, possibly consisting of two blended
components. The evolution with frequency is surprising. The profile
widens between 0.69 and 1.4~GHz before narrowing again at 3.1~GHz.
The circular polarization swings from negative to positive
across the pulse at all frequencies and it would appear likely that
the midpoint of the swing gives $\phi_0$. The linear polarization is
low and the behaviour of the PA swing is rather unclear. It is difficult
to assign PA$_0$ with any confidence.

\noindent{\bf PSR~J1801$-$2451}: This pulsar has a small characteristic
age of only 15~kyr. It has an associated pulsar wind nebula which
is located at the tip of the supernova remnant SNR G5.4$-$1.2 (the `Duck').
Arguments have been put forward associating the pulsar with the
SNR; this either requires a high proper motion or a rather old age
\cite{gf00}. A VLBI measurment of the proper motion appears to
sway the balance against the association \cite{tbg02}. The detection of
an X-ray tail by Kaspi et al. (2001)\nocite{kggl01} would appear to
show the pulsar moving almost due west albeit at a rather low
velocity \cite{bgc+06}.

At low frequencies, the profile is highly scattered but at frequencies
above $\sim$1~GHz, the profile consists of a narrow, highly polarized
component typical of other young pulsars. We have determined an RM
to this pulsar for the first time.  The swing of PA is moderately
steep. If we associate the peak of the profile with $\phi_0$ then
PA$_0$ is $-$55\degr.  However, we cannot be certain about this
identification because the profile may simply be a leading or trailing
edge (like PSRs J0543+2329 above) with the magnetic crossing not seen.

\noindent{\bf PSR~J1820$-$0427}: At low frequencies the pulse profile
consists of a single, narrow component with moderate linear and
negative circular polarization. Rapid evolution of the profile occurs
with frequency: at 3.1~GHz a strong leading component dominates and
there is also a hint of a trailing component. The leading component
shows higher linear polarization. The PA features an orthogonal jump
near the leading edge of the profile followed by a steep rise through
the centre of the pulse. The RVM fit is reasonably good, with an
accurate determination of $\phi_0$. At this location, PA$_0$ is 42\degr.

\noindent{\bf PSR~J1825$-$0935}: This pulsar is interesting for a variety
of reasons and its characteristics have been discussed extensively
in Gil et al. (1994)\nocite{gjk+94}. In brief the profile consists
of a small leading component and a dominant trailing component, 
followed almost 180\degr\ later by an `interpulse' component.
The leading component is highly
linearly polarized whereas the dominant component has linear
polarization which declines at higher frequencies. The `interpulse'
component has no polarization. There is little change in component
separation as a function of frequency. Remarkably, the interpulse and
the leading component appear to exchange information (Gil et al. 1994)
which is difficult to understand in current models of the radio emission.
Speculation continues as to whether we
see emission from both poles in an orthogonal rotator or from a wide beam 
from a single pole. Most recently, Dyks et al. (2005)\nocite{dzg05} speculate
that the leading and trailing main pulse components arise from conventional
emission within a standard magnetosphere. The interpulse, however, is
seen when the emission from the leading component is reversed and propagates
inwards rather than outwards. At a stroke, this interesting idea 
explains both the component separations and the common modulation properties.

We have made an attempt at an RVM fit to our profile (using only the main
pulse as the interpulse is unpolarized). With the addition of an orthogonal
jump between the two components of the main pulse, we find a plausible solution
which yields $\alpha\sim$95\degr, $\beta\sim -7$\degr. In this solution
the RVM inflexion point trails the profile midpoint by 2.2\degr, and
PA$_0$ is 10\degr.  Although these numbers should be treated with caution
it is interesting that $\alpha+\beta$ is close to 90\degr, as predicted
in the Dyks et al. (2005) model.

\noindent{\bf PSR~J1841$-$0425}: The pulse profile is highly scattered at
frequencies below $\sim$1~GHz. At 1.4 and 3.1~GHz the profile consists
of a single, highly polarized component, strikingly similar to
PSR~1801$-$2451 above. If we identify the profile midpoint with $\phi_0$
then PA$_0$ is 3\degr.

\noindent{\bf PSR~J1850+1335}: The profile of this pulsar does not
vary greatly with frequency. It consists of a single, moderately
polarized component with a degree of negative circular
polarization. The 1.4~GHz profile in Weisberg et al. (1999)\nocite{wcl+99},
with higher signal to noise shows the profile extends to later longitudes.
The PA swing declines gently throughout. At the profile
peak, the PA is $-$45\degr. Again, however, if we are only seeing a partial
cone, this may not reflect PA$_0$.
Note that in the 0.4~GHz profile of Gould
\& Lyne (1998)\nocite{gl98} the profile rather curiously appears to be
unpolarized.

\noindent{\bf PSR~J1915+1009}: The profile consists of a single component
at all frequencies with moderate amounts of both linear and circular
polarization. The PA swing is entirely flat at low frequencies but has
an orthogonal mode jump at 3.1~GHz. If we assign the profile peak
as $\phi_0$ then PA$_0$ is 85\degr.

\noindent{\bf PSR~J1916+0951}: This pulsar has a double peaked profile
with both components of roughly equal strengths at 1.4~GHz. At lower
frequencies the leading component dominates whereas at higher frequencies
the opposite is true \cite{kkwj98}. The linear polarization is confined to the
leading component.  Weisberg et al. (1999)\nocite{wcl+99} classify
the leading component as the core with the leading conal component not
seen.  The PA swing is not very revealing and determination of $\phi_0$
is too ambiguous to derive PA$_0$.

\noindent{\bf PSR~J1937+2544}: The profile of this pulsar consists of
two equal components and is relatively wide, covering more than
30\degr\ of longitude. The leading component is highly polarized
unlike the trailing component. The trailing component starts to
dominate at frequencies above $\sim$3~GHz. The saddle region between
the components appears to fill in at high frequencies, a behaviour
opposite to that seen in many other double profiles. The PA swing is
flat across the first component and steeper across the trailing
component (see also Weisberg et al. 1999).
Additionally, there is a late orthogonal jump at 0.67~GHz.
RVM fits to the profile are good, yielding PA$_0$ of $-$9\degr\
at the location of the inflexion point.

\noindent{\bf PSR~J2048$-$1616}: The PA swing in this pulsar shows the classic
RVM response at all three of our observing frequencies. A good fit
can be obtained and $\phi_0$ can be well determined and lies exactly at
the midpoint between the two outer components. The value of PA$_0$
is $-$13\degr, similar to that obtained by Rankin (2007).
Of the three components, the inner one has 
the steeper spectral index. The inner component also
lags both $\phi_0$ and the midpoint of the outer components.

\noindent{\bf PSR~J2330$-$2005}: Two clear components are seen in the
profile, with possibly a hint of a weak, third trailing component
(see also Gould \& Lyne 1998\nocite{gl98})
There is little variation in either the profile
shape, width or polarization characteristics with frequency although
the overall level of linear polarization declines slowly with
increasing frequency. The PA swing is peculiar with a flat PA on the wings
and a steep, complicated PA swing through the symmetrical centre of the
profile. One possible interpretation is that this is a 180\degr\ swing of PA
as the sightline passes very close to the pole. In fact, the value
of PA at 0.67~GHz in the pulse centre is almost exactly 90\degr\ away
from the average of the PA in the wings. We therefore assign a
PA$_0$ of 21\degr\ for this pulsar.

\begin{figure*}
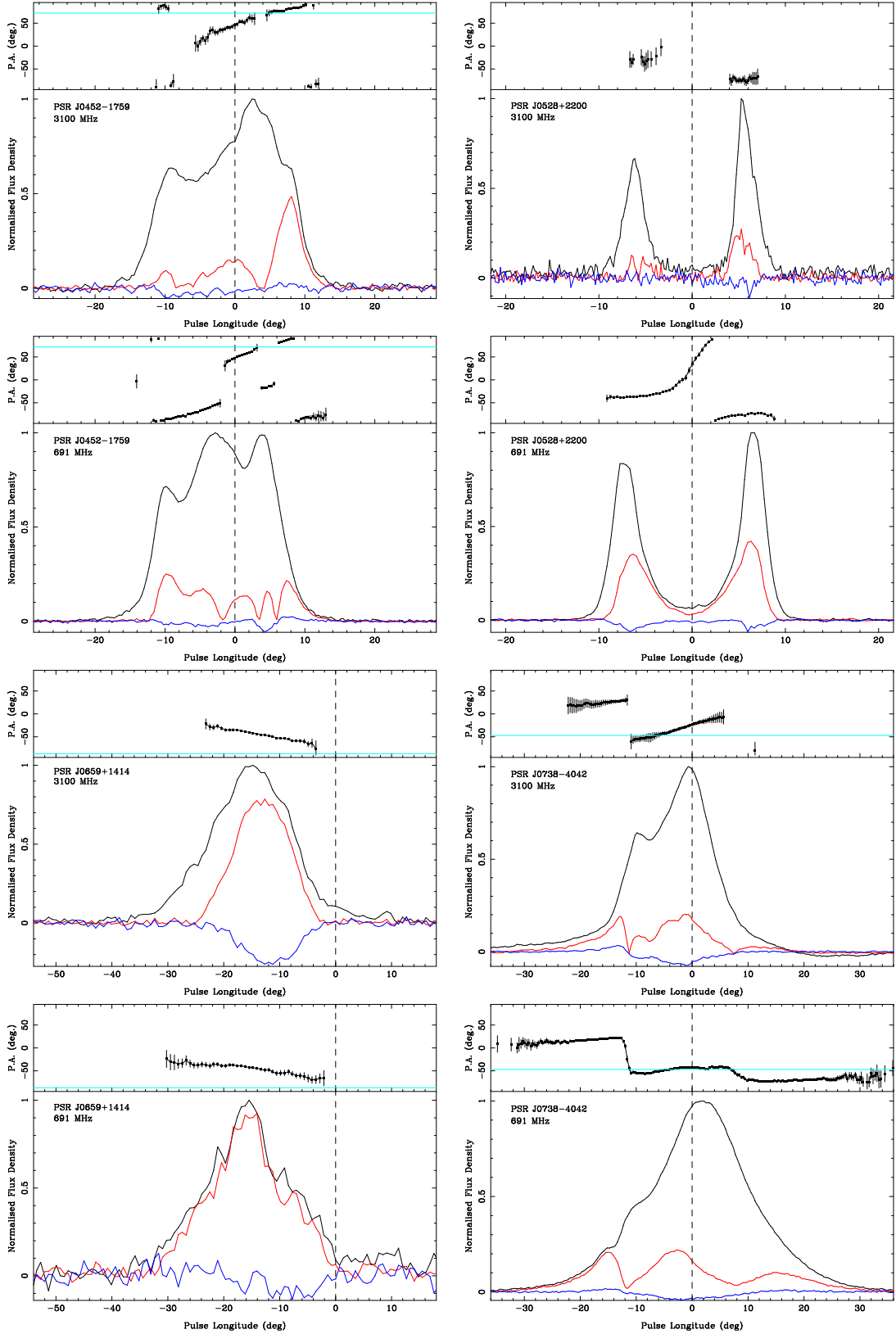

\begin{tabular}{cc}
\psfig{figure=fig1a.ps,angle=-90,width=7cm} &
\psfig{figure=fig1b.ps,angle=-90,width=7cm} \\
\psfig{figure=fig1c.ps,angle=-90,width=7cm} &
\psfig{figure=fig1d.ps,angle=-90,width=7cm} \\
\psfig{figure=fig1e.ps,angle=-90,width=7cm} &
\psfig{figure=fig1f.ps,angle=-90,width=7cm} \\
\psfig{figure=fig1g.ps,angle=-90,width=7cm} &
\psfig{figure=fig1h.ps,angle=-90,width=7cm} \\
\end{tabular}
\caption{(a)-(h). Polarization profiles at 0.69 and 3.1~GHz for
PSR~J0452$-$1759 (top left),
PSR~J0528+2200 (top right), PSR~J0659+1414 (bottom left) and
PSR~J0738$-$4042 (bottom right).
On each plot, the bottom panel shows the total intensityi (black, thick
line), linear (grey line) and circular (thin, grey line)
polarizations. The top panel shows the PA of the polarized radiation
at the pulsar. PA$_v$ (where known) is shown as a solid horizontal line.
Pulse longitude 0.0 denotes the closest approach of the line of sight
to the magnetic pole (dashed vertical line).
The PA at this longitude gives PA$_0$ as described in the text.}
\end{figure*}
\begin{figure*}
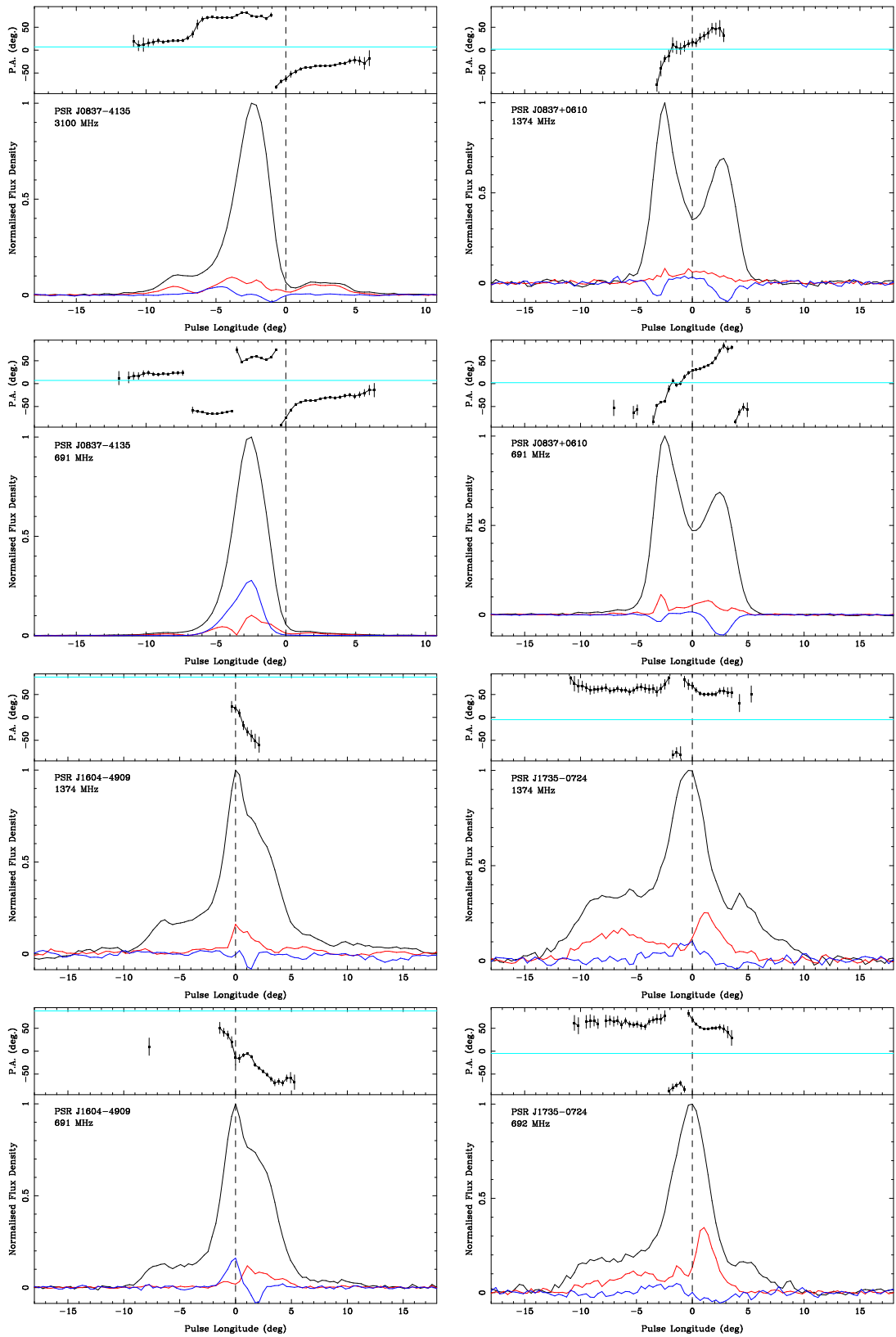

\begin{tabular}{cc}
\psfig{figure=fig2a.ps,angle=-90,width=7cm} &
\psfig{figure=fig2b.ps,angle=-90,width=7cm} \\
\psfig{figure=fig2c.ps,angle=-90,width=7cm} &
\psfig{figure=fig2d.ps,angle=-90,width=7cm} \\
\psfig{figure=fig2e.ps,angle=-90,width=7cm} &
\psfig{figure=fig2f.ps,angle=-90,width=7cm} \\
\psfig{figure=fig2g.ps,angle=-90,width=7cm} &
\psfig{figure=fig2h.ps,angle=-90,width=7cm} \\
\end{tabular}
\caption{(a)-(h). Polarization profiles at 0.69 and 3.1~GHz for
PSR~J0837$-$4135 (top left)
and at 1.4 and 3.1~GHz for PSR~J0837+0610 (top right)
PSR~J0837$-$4135 (top right), PSR~J1604$-$4909 (bottom left) and
PSR~J1735$-$0724 (bottom right) 
See Fig.~1 for details.}
\end{figure*}
\begin{figure*}
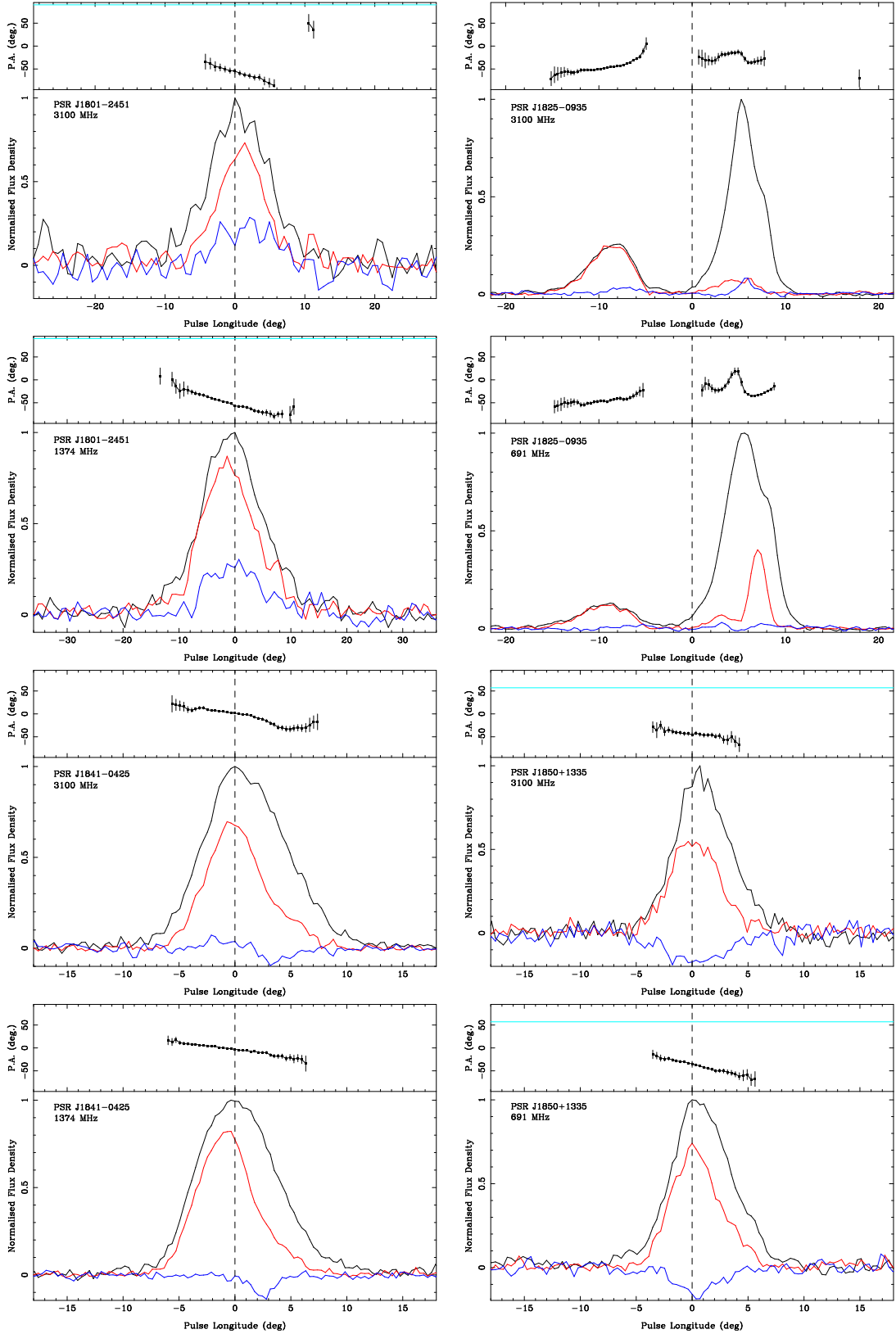

\begin{tabular}{cc}
\psfig{figure=fig3a.ps,angle=-90,width=7cm} &
\psfig{figure=fig3b.ps,angle=-90,width=7cm} \\
\psfig{figure=fig3c.ps,angle=-90,width=7cm} &
\psfig{figure=fig3d.ps,angle=-90,width=7cm} \\
\psfig{figure=fig3e.ps,angle=-90,width=7cm} &
\psfig{figure=fig3f.ps,angle=-90,width=7cm} \\
\psfig{figure=fig3g.ps,angle=-90,width=7cm} &
\psfig{figure=fig3h.ps,angle=-90,width=7cm} \\
\end{tabular}
\caption{(a)-(h). Polarization profiles at 1.4 and 3.1~GHz for
PSR~J1801$-$2451 (top left) and PSR~J1841$-$0425 (bottom left)
and at 0.69 and 3.1~GHz for PSR~J1825$-$0935 (main pulse only; top right) and
PSR~J1850+1335 (bottom right). See Fig.~1 for details.}
\end{figure*}
\begin{figure*}
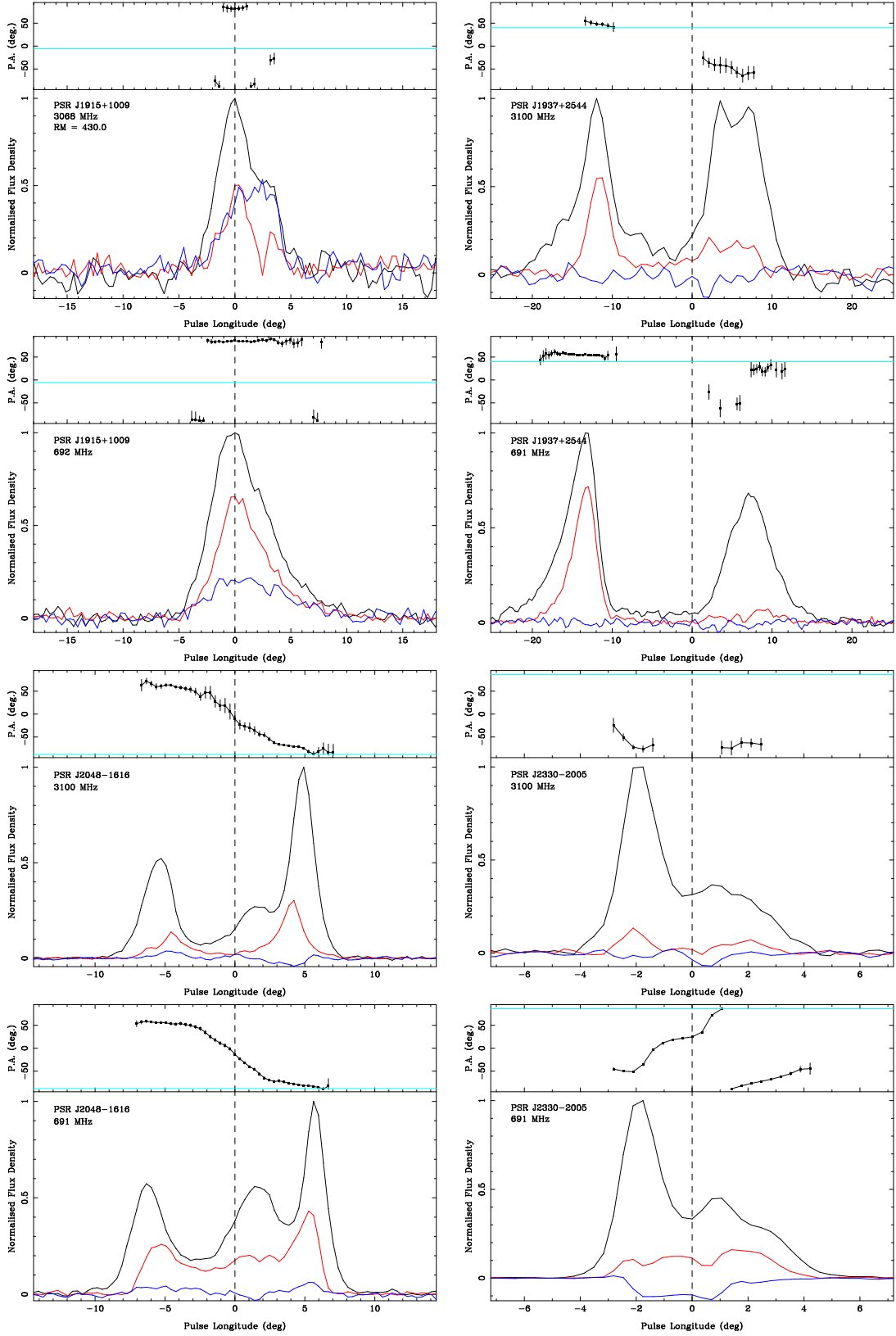

\begin{tabular}{cc}
\psfig{figure=fig4a.ps,angle=-90,width=7cm} &
\psfig{figure=fig4b.ps,angle=-90,width=7cm} \\
\psfig{figure=fig4c.ps,angle=-90,width=7cm} &
\psfig{figure=fig4d.ps,angle=-90,width=7cm} \\
\psfig{figure=fig4e.ps,angle=-90,width=7cm} &
\psfig{figure=fig4f.ps,angle=-90,width=7cm} \\
\psfig{figure=fig4g.ps,angle=-90,width=7cm} &
\psfig{figure=fig4h.ps,angle=-90,width=7cm} \\
\end{tabular}
\caption{(a)-(h). Polarization profiles at 0.69 and 3.1~GHz for
PSR~J1915+1009 (top left) and PSR~J1937+2544 (top right),
PSR~J2048$-$1616 (bottom left) and
PSR~J2330$-$2005 (bottom right). See Fig.~1 for details.}
\end{figure*}
\begin{figure*}
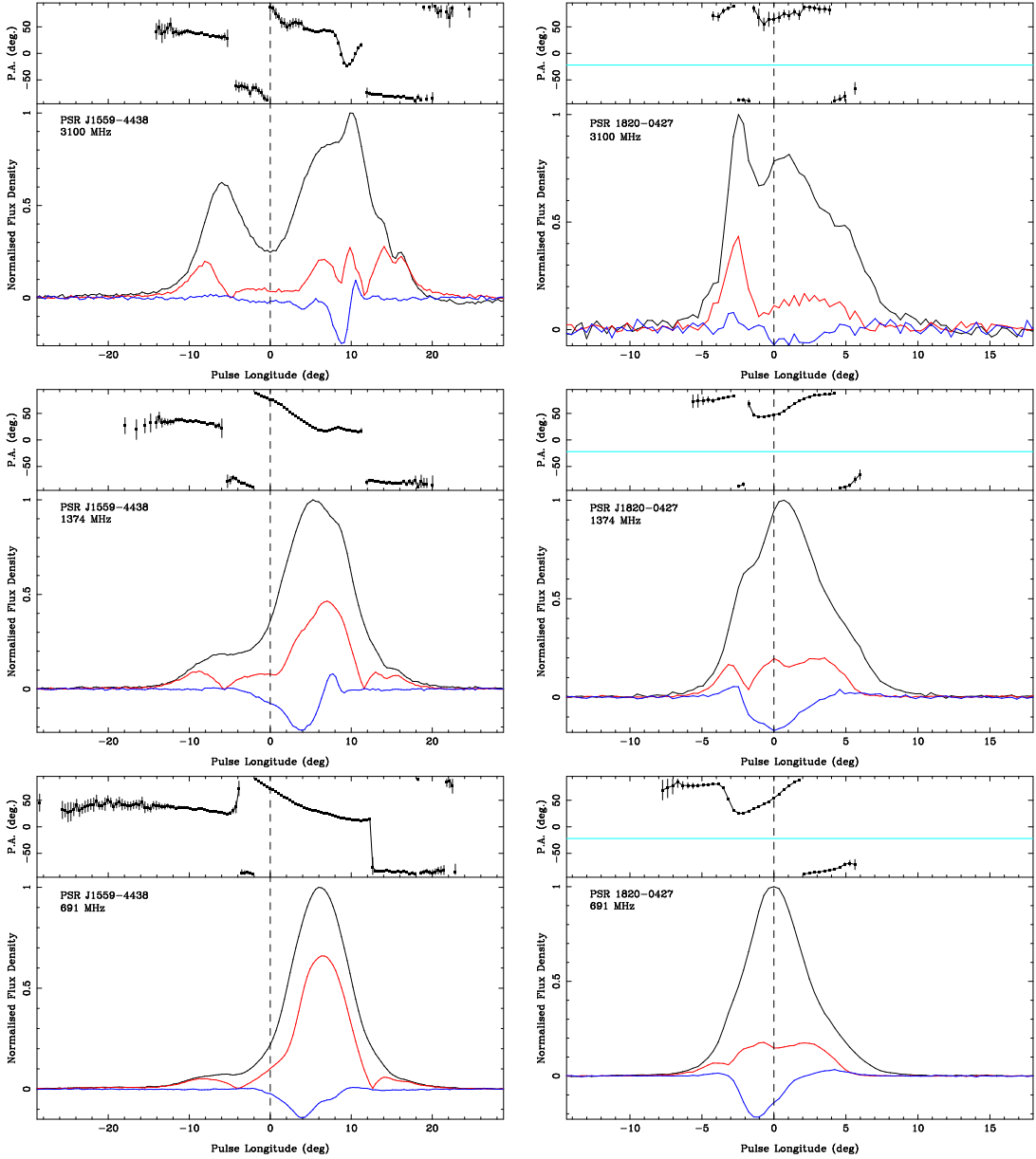

\begin{tabular}{cc}
\psfig{figure=fig5a.ps,angle=-90,width=7cm} &
\psfig{figure=fig5b.ps,angle=-90,width=7cm} \\
\psfig{figure=fig5c.ps,angle=-90,width=7cm} &
\psfig{figure=fig5d.ps,angle=-90,width=7cm} \\
\psfig{figure=fig5e.ps,angle=-90,width=7cm} &
\psfig{figure=fig5f.ps,angle=-90,width=7cm} \\
\end{tabular}
\caption{(a)-(h). Polarization profiles at 0.69, 1.4 and 3.1~GHz for
PSR~J1559$-$4438 (top left),
PSR~J1820$-$0427 (bottom right). See Fig.~1 for details.}
\end{figure*}
\begin{figure*}
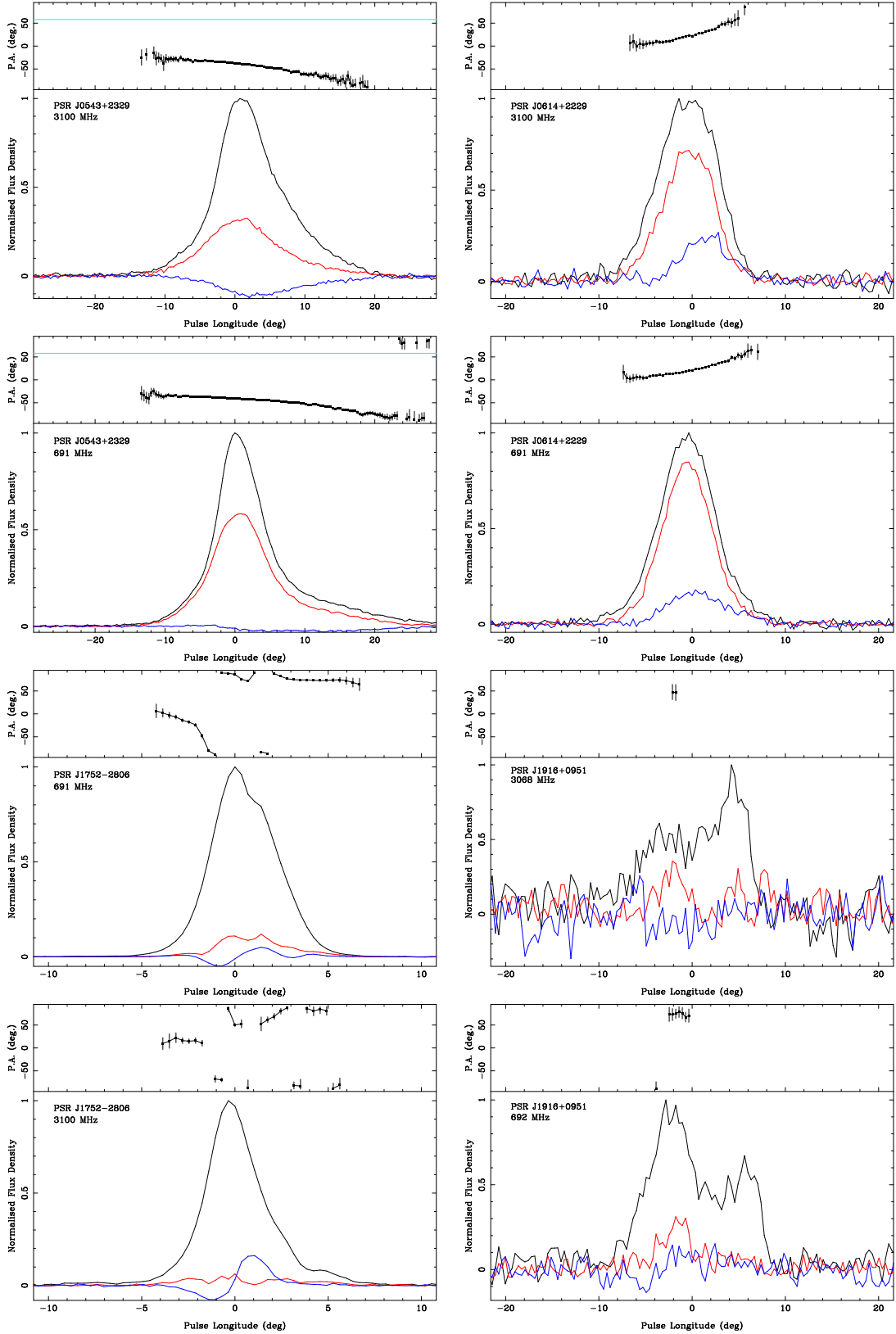

\begin{tabular}{cc}
\psfig{figure=fig6a.ps,angle=-90,width=7cm} &
\psfig{figure=fig6b.ps,angle=-90,width=7cm} \\
\psfig{figure=fig6c.ps,angle=-90,width=7cm} &
\psfig{figure=fig6d.ps,angle=-90,width=7cm} \\
\psfig{figure=fig6e.ps,angle=-90,width=7cm} &
\psfig{figure=fig6f.ps,angle=-90,width=7cm} \\
\psfig{figure=fig6g.ps,angle=-90,width=7cm} &
\psfig{figure=fig6h.ps,angle=-90,width=7cm} \\
\end{tabular}
\caption{(a)-(d). Polarization profiles at 0.69 and 3.1~GHz for
PSR~J0543+2329 (top left), PSR~J0614+2229 (top right),
PSR~J1752$-$2806 (bottom left) and PSR~J1916+0951 (bottom right).
PA$_0$ could not be assigned to these pulsars and so the longitude
scale is arbitrary. See Fig.~1 for other details.}
\end{figure*}
\subsection{Emission Heights and Time Alignment}
\noindent{\bf PSR~J0452$-$1759}: The excellent RVM fit 
allows a determination of both the geometrical and A/R heights in
this pulsar. The inflexion point of the RVM lags the midpoint of
the profile by $\sim$2\degr and this implies an emission height of
$\sim$200~km. In contrast, the observed pulse width and geometrical
information yields an emission height for the outer edges at $\sim$700~km.
The time alignment for the pulsar shows alignment on the rising edge
and that the emerging component at 3.1~GHz occurs later than emission
at 0.69~GHz.

\noindent{\bf PSR~J0528+2200}: The inflexion point of the RVM lies
(within the errors) at the symmetry point of the pulse profile.  We
can use this information to derive an upper limit on the A/R emission
height of 200~km.  Furthermore, the time alignment shows that the
midpoint of the 0.69 and 3.1~GHz profiles lie at the same longitude
implying a very small difference between the emission heights at the
two frequencies at least in the pulse centre.  These results are
reasonably consistent with the delay heights derived by Mitra \& Li
(2004)\nocite{ml04}.

\noindent{\bf PSRs~J0543+2329, J0614+2229, J0659+1414, J1841$-$0425, J1850+1335
and J2330$-$2005}: The time 
alignment shows an identical location for the peak of the 
0.69 and 3.1~GHz profiles in these pulsars.
Although we cannot place an absolute emission height
on the emission, the difference in the height between the two
frequencies must nevertheless be small because of the excellent agreement
between the two profiles and lack of any A/R offset.

\noindent{\bf PSR~J0738$-$4042}: The RVM fit to this pulsar is excellent
at 1.4~GHz which allows a determination of the emission height through
the pulse width. Using $\alpha$ of 62\degr\ and $\beta$ of 13\degr\
gives an emission height of 900~km. Furthermore the profile mid point
(and the centre of the two emission peaks) occur some 4\degr\ earlier than
the RVM inflexion point. This yields an A/R height determination of 300~km.
The alignment of this pulsar was considered in 
Karastergiou \& Johnston (2006)\nocite{kj06}.  They showed that,
if the alignment was made on the PA swing then the leading orthogonal
jump occurred earlier at lower frequencies. This is confirmed by
the time alignment here. The leading orthogonal jump occurs
early at 0.69~GHz by about 4\degr.

\noindent{\bf PSR~J0837$-$4135}: There is almost a perfect alignment
between the profiles at the two frequencies. On the other hand
the inflexion point of the RVM occurs later than the midpoint
as expected in the BCW model. This implies an emission height of $\sim$400~km
but a difference in emission height of only 60~km between the
two frequencies.

\noindent{\bf PSR~J0837+0610}: The profile widths at 0.69 and 3.1 GHz
are very similar although the higher frequency profile has a much
steeper falling edge than the low frequency one. There is also more of
a saddle between the two emission components at the lower frequency.
The time alignment shows the profile perfectly aligned on the 10\%
power points. The apparent shift in the peak of the trailing component
is likely simply due to the weakening of the central, blended
component at high frequencies.  Although the PA swing is somewhat
complex at low frequencies, the steepest part of the swing trails the
symmetry point of the profile and leads to a height determination of
$\sim$400~km.

\noindent{\bf PSR~J1559$-$4438}: The RVM fit is excellent for this
pulsar at lower frequencies. The longitude of the inflexion point is
offset from the profile centre by less than $\sim$2\degr. This implies
an emission height of under 100~km, at least for the core region.
However, we can also determine the emission height geometrically for
this pulsar because $\alpha$, $\beta$ and the pulse width are well determined.
Assuming emission extends out to the last open field line, the emission
height at the wings is then $\sim$650~km. This is evidence that the
emission height appears to decline from the outside to the centre of
pulse profiles \cite{gg03}.
The profile evolution with frequency is hard to understand.
The time alignment would indicate that the trailing peaks
align and yet at the high frequencies the trailing edge seems to
split into at least four different components.

\noindent{\bf PSRs~J1604$-$4909 and J1752$-$2806}: The profile widths 
at 0.69 and 3.1~GHz are very similar in these two pulsars 
even though the outrider components become relatively
brighter at the higher frequency. The time alignment is as expected
with a perfect match between the phase of the central peaks.
The PA swings are too complex to allow emission heights to be derived.

\noindent{\bf PSR~J1820$-$0427}: The low frequency profile shows a
curiously featureless single component whereas the profile at high
frequency shows at least three components. The rather steep PA swing
at low frequencies has become significantly flatter at the high
frequency. The inflexion point is coincident with the peak of the
profile at low frequencies implying a low emission height.

\noindent{\bf PSR~J1825$-$0935}: If the RVM fit we outlined in the
previous section is correct, then the 2.2\degr\ lag between the RVM
inflexion point and the profile midpoint implies an emisison height
of 350~km. There is little change in the profile width between 
0.69 and 3.1~GHz, implying only a small height difference in
emission heights.

\noindent{\bf PSR~J1937+2544}: In spite of the somewhat sparse PA data
points, a good RVM fit can be obtained to the data at 0.69 and 3.1~GHz
after allowing for an orthogonal jump at the lower frequency.
These fits show that the inflexion point of the RVM is delayed with respect
to the profile centre by 4\degr. This implies an emission height using
A/R of 170~km with little difference between the high and low frequencies.
Geometrical heights can be estimated from the profile width
and the RVM values. The higher frequency profile is narrower than the 
low frequency profile, leading to height estimates of 200 and 250~km
respectively.

\noindent{\bf PSR~J2048$-$1616}: Recent attempts have been made to
classify this pulsar and determine emission heights of the various
components \cite{gg03,ml04}. Both these papers claim that the central
component clearly visible at low frequencies is the core in spite of
the fact that it actually {\it lags} the inflexion point of the RVM.
We disagree with this interpretation finding it more likely that this
component is an inner conal component.

We can use the excellent RVM fit at 0.69~GHz (yielding $\alpha$ of 34\degr\
and $\beta$ of $-$1.6\degr) and observed width to compute a geometrical
height of 300~km. Furthermore we can use the method of Gangadhara \&
Gupta (2001)\nocite{gg01} to compute the A/R heights. Here though we
use the RVM inflexion point as the core location rather than the offset
central component as used by others. The difference in the separation 
between the outer peaks and the RVM inflexion point is 0.3\degr\ at 
3.1~GHz and 0.4\degr\ at 0.69~GHz yielding emission heights of 240 and
330~km respectively in excellent agreement with the geometrical height.
%
%Finally, the time alignment shows that the 0.69~GHz profile has been
%retarded by 0.7\degr\ with respect to the alignment of the RVM curves.

\section{Discussion}
\subsection{Emission Heights}
\begin{table}
\caption{Emission heights for 9 pulsars.}
\begin{tabular}{llrr}
\hline & \vspace{-3mm} \\
& & \multicolumn{1}{c}{h$_{\rm A/R}$} & \multicolumn{1}{c}{h$_{\rm GEO}$}\\
Jname & Bname & \multicolumn{1}{c}{(km)} & \multicolumn{1}{c}{(km)}\\
\hline & \vspace{-3mm} \\
J0452$-$1759 & B0450$-$18 & 200 & 700\\
J0528+2200 & B0525+21 & $<$200 \\
J0738$-$4042 & B0736$-$40 & 300 & 900 \\
J0837+0610 & B0834+06  & 400\\
J0837$-$4135 & B0835$-$41 & 400\\
J1559$-$4438 & B1556$-$44 & $<$100 & 650 \\
J1825$-$0935 & B1822$-$09 & 350 \\
J1937+2544 &  B1935+25 & 170 & 250 \\
J2048$-$1616 & B2045$-$16 & 330 & 300\\
\hline & \vspace{-3mm} \\
\end{tabular}
\label{emh}
\end{table}

\begin{table*}
\caption{PA$_v$, PA$_0$ and their difference, $\Psi$, for our current sample.
The figure in brackets gives the error in the last digit(s).}
\begin{tabular}{llccrrr}
\hline & \vspace{-3mm} \\
\multicolumn{1}{c}{Jname} & \multicolumn{1}{c}{Bname} 
& \multicolumn{1}{c}{log[age]} & \multicolumn{1}{c}{V$_T$}
& \multicolumn{1}{c}{PA$_v$} & \multicolumn{1}{c}{PA$_0$}
& \multicolumn{1}{c}{$\Psi$} \\
& & \multicolumn{1}{c}{(yr)} & \multicolumn{1}{c}{kms$^{-1}$}
& \multicolumn{1}{c}{(deg)} & \multicolumn{1}{c}{(deg)}
& \multicolumn{1}{c}{(deg)}\\
\hline & \vspace{-3mm} \\
J0452$-$1759 & B0450$-$18 & 6.2 & 185 & 72(23) & 47(3) & 25(23) \\
J0659+1414   & B0656+14   & 5.0 & 65  & 93.1(4) & $-$86(2) & $-$1(5) \\
J0738$-$4042 & B0736$-$40 & 6.6 & 180 & 313(5) & $-$21(2) & $-$26(5) \\
J0837+0610   &  B0834+06  & 6.5 & 170 & 2(5) & 18(5) & $-$16(7) \\
J0837$-$4135 & B0835$-$41 & 6.5 & 360 & 187(6) & $-$84(5) & $-$89(8) \\
J1604$-$4909 & B1600$-$49 & 6.7 & 510 & 268(6) & $-$17(3) & $-$75(7) \\
J1735$-$0724 & B1732$-$07 & 6.7 & 570 & 355(3) & 55(5) & $-$60(6) \\
J1801$-$2451 & B1757$-$24 & 4.2 & 300 & 270 & $-$55(5) & $-$35(5) \\
J1820$-$0427 & B1818$-$04 & 6.2 & 190 & 338(17) & 42(3) & $-$64(17) \\
J1850+1335   & B1848+13   & 6.6 & 300 & 237(16) & $-$45(3) & $-$78(16) \\
J1915+1009   & B1913+10   & 5.6 & 280 & 174(15) & 85(3) & 89(15)\\
J1937+2544   &  B1935+25  & 6.7 & 210 & 220(9) & $-$9(5) & 49(10) \\
J2048$-$1616 & B2045$-$16 & 6.5 & 330 & 92(2) & $-$13(5) & $-$75(6) \\
J2330$-$2005 & B2327$-$20 & 6.7 & 180 & 86(2) & 21(10) & 65(10) \\
\hline & \vspace{-3mm} \\
\end{tabular}
\label{results}
\end{table*}

We have computed emission heights for 9 of the pulsars in our sample.
These are listed in the Table~\ref{emh}. The last two columns show the
measured heights from the A/R method and the geometrical method.  The
A/R technique measures the height of the central component, as it uses
information from the offset between the central location and the location
of the inflexion point of the RVM. The geometrical height is derived
from the outer limits of the profile under the assumption that the
emission extends to the last open field lines. Gangadhara \& Gupta
(2003) have shown that this may not be the case and measure the ratio of
the radius of last emitting field line to the radius of the polar cap,
$s$, to be $\sim$0.7. The emission heights derived here then would
need to be divided by $s^2$ (i.e. would increase by factor of
$\sim$2).  In any case it is interesting that our results show
that the heights of the central emitting regions
are significantly lower than those of the outer cones. We have also
seen that between 0.69 and 3.1~GHz the emission heights do not greatly
change, as expected using the radius to frequency mapping relationship
of e.g. Thorsett (1991)\nocite{tho91a}.

The results are in line with a recent model proposed by
Karastergiou \& Johnston (2007)\nocite{kj07}. In that model, all radio
emission arises from near the last open field lines, with a maximum
emission height around 1000~km. Different emission heights then cause
the observed pulse profile with emission from higher in the magnetosphere
associated with the outer parts of the profile. This is entirely consistent
with the results for the central and outer components presented here.

\subsection{Rotation and velocity alignment}
In Table~\ref{results} we summarise our results
for those pulsars with errors less than 25\degr. 
Column 4 shows the traverse velocity of the pulsar
using the proper motion and distance values from Table~\ref{sources}.
Columns 5 and 6 show
the position angles of the velocity vector, PA$_v$, and of the linear
polarisation, PA$_0$, at $\phi_0$, the closest approach of the
line-of-sight to the magnetic pole (see Equation 1), taken from
Table~\ref{sources}.
Column 6 shows the offset, $|\Psi|$, between PA$_v$ and PA$_0$ where
$\Psi$ is defined as
\begin{equation}
\Psi = {\rm PA}_v - {\rm PA}_0 ; \,\,\,\,\, -90\degr \leq \Psi \leq 90\degr
\end{equation}
We force $\Psi$ to lie between $-$90\degr\ and 90\degr by rotating
PA$_0$ by $\pm$180\degr\ as necessary (note that any polarized PA has
by nature a 180\degr\ ambiguity).
The error in $\Psi$ comes from a quadrature sum of the errors in
PA$_v$ and PA$_0$.

Of the sample, 7 pulsars appear to have the velocity and rotation vectors
plausibly correlated (PSRs J0659+1414, J0837+0610, J0837$-$4135,
J1604$-$4909, J1850+1335, J1915+1009 and J2048$-$1616).
Of these 7, all but PSRs~J0659+1414 and J0837+0610 show
the difference in the vectors to be 90\degr\ rather than 0\degr.
As we argued in Johnston et al. (2005) we believe this implies that
the velocity and rotation vectors are actually aligned and that the radio 
emission predominantly radiates perpendicular to the magnetic field lines. 
Table~\ref{results} shows no obvious correlation between the pulsar's
velocity and $\psi$. The slowest moving pulsar in the sample,
PSR~J0659+1414 appears to be aligned whereas the fast moving
PSR~J1735$-$0724 is not (although it is also somewhat older).
The possible misalignment in the angles for PSR~J1801$-$2401
is particularly interesting in light of its probable
association with the `Duck' SNR, although as cautioned in the section
above, the location of PA$_0$ is not watertight in this pulsar.

The correlation between the velocity and rotation vectors therefore appears
less strong than in Johnston et al. (2005) and Rankin (2007).
However, the pulsars in the current sample are both older and more distant
than the earlier sample. Both these parameters make it more likely that 
the gravitational pull of the Galaxy has altered the pulsar's proper
motion vector from its birth direction.
Although the numbers are small and the uncertainties large,
the results do not obviously follow the prediction of 
Ng \& Romani (2007) who postulate
that only the fast moving pulsars should show a correlation 
between their velocity and rotation vectors.
In a further paper, we will investigate in detail the properties of
the pulsars which clearly show this correlation, and distinguish them
from the pulsars that do not (Carr et al., In Preparation).

\section{Conclusions}
We have presented polarization observations of 22 pulsars.
We have compared the orientation of the spin and velocity vectors 
and find that, for the 14 pulsars for which we
were able to determine both vectors, that 7 are plausibly aligned,
a fraction which is lower than, but consistent with, earlier measurements.
There appears to be no obvious correlation between a pulsar's velocity
and its alignment angle.
We have used profiles obtained simultaneously
at widely spaced frequencies to compute the radio emission heights of
the pulsars in our sample. All have emission heights less than $\sim$1000~km
and show that radiation from the centre of the profile occurs from lower
in the magnetosphere than that from the outer parts of the profile.

\section*{Acknowledgments}
The authors would like to thank the referee, Joanna Rankin for her
careful reading of this manuscript and the helpful suggestions
for improving the paper.
The Australia Telescope is funded by the Commonwealth of 
Australia for operation as a National Facility managed by the CSIRO.
AK acknowledges financial support from the 6th European Community
Framework programme through a Marie Curie, Intra-European Fellowship.

\bibliography{modrefs,psrrefs,crossrefs}
\bibliographystyle{mn}
\label{lastpage}
\end{document}